\documentclass[amsmath,showpacs,nofootinbib,11pt,superscriptaddress]{revtex4-2}
\usepackage{graphicx}
\usepackage{dcolumn}
\usepackage{bm}
\usepackage{subcaption}
\usepackage{color} 
\usepackage{slashed}
\usepackage{float}
\usepackage{amsfonts}
\usepackage{multirow}
\usepackage{lineno}
\usepackage[colorlinks=true,linkcolor=blue,citecolor=blue,urlcolor=blue]{hyperref}

\begin{document}
\newcommand{\overlr}{\stackrel{\leftrightarrow}{\partial}}
\newcommand{\hs}{\hspace*{0.5cm}}
\newcommand{\vs}{\vspace*{0.5cm}}
\newcommand{\be}{\begin{equation}}
\newcommand{\ee}{\end{equation}}
\newcommand{\bea}{\begin{eqnarray}}
\newcommand{\eea}{\end{eqnarray}}
\newcommand{\beal}{\begin{align}}
\newcommand{\eeal}{\end{align}}
\newcommand{\ben}{\begin{enumerate}}
\newcommand{\een}{\end{enumerate}}
\newcommand{\bde}{\begin{widetext}}
\newcommand{\ede}{\end{widetext}}
\newcommand{\nn}{\nonumber}
\newcommand{\crn}{\nonumber \\}
\newcommand{\Tr}{\mathrm{Tr}}
\newcommand{\non}{\nonumber}
\newcommand{\noi}{\noindent}
\newcommand{\al}{\alpha}
\newcommand{\la}{\lambda}
\newcommand{\bet}{\beta}
\newcommand{\ga}{\gamma}
\newcommand{\va}{\varphi}
\newcommand{\om}{\omega}
\newcommand{\pa}{\partial}
\newcommand{\+}{\dagger}
\newcommand{\fr}{\frac}
\newcommand{\bc}{\begin{center}}
\newcommand{\ec}{\end{center}}
\newcommand{\Ga}{\Gamma}
\newcommand{\de}{\delta}
\newcommand{\De}{\Delta}
\newcommand{\ep}{\epsilon}
\newcommand{\varep}{\varepsilon}
\newcommand{\ka}{\kappa}
\newcommand{\La}{\Lambda}
\newcommand{\si}{\sigma}
\newcommand{\Si}{\Sigma}
\newcommand{\ta}{\tau}
\newcommand{\up}{\upsilon}
\newcommand{\Up}{\Upsilon}
\newcommand{\ze}{\zeta}
\newcommand{\ps}{\psi}
\newcommand{\Ps}{\Psi}
\newcommand{\ph}{\phi}
\newcommand{\vph}{\varphi}
\newcommand{\Ph}{\Phi}
\newcommand{\Om}{\Omega}

\newcommand{\ITAR}{Institute of Theoretical and Applied Research, Duy Tan University, Hanoi 10000, Vietnam}
\newcommand{\IAST}{Laboratory of Advanced Materials and Natural Resources, Institute for
	Advanced Study in Technology, Ton Duc Thang University, Ho Chi Minh City,
	Vietnam}
\newcommand{\DAS}{Faculty of Applied Sciences, Ton Duc Thang University, Ho Chi Minh City,
	Vietnam}
\newcommand{\IOP}{Institute of Physics,  Vietnam Academy of Science and Technology, 10 Dao Tan, Ba Dinh, Hanoi 10000, Vietnam}
\newcommand{\SPtwo}{Department of Physics, Hanoi Padegogical 2, Xuan Hoa, Phu Tho, Vietnam}
\newcommand{\VanlangA}{Subatomic Physics Research Group, Science and Technology Advanced Institute, Van Lang University, Ho Chi Minh City 70000, Vietnam}
\newcommand{\VanlangB}{Faculty of Technology, Van Lang University, Ho Chi Minh City 70000, Vietnam} 
\newcommand{\AdrHEPC}{Phenikaa Institute for Advanced Study, Phenikaa University, Duong Noi, Hanoi 100000, Vietnam}


\title{Probing muon anomaly and lepton flavor violation with scalar leptoquarks in the 331LHN model}

\author{D. T. Binh}
\email{dinhthanhbinh@tdtu.edu.vn}
\affiliation{\IAST}
\affiliation{\DAS}

\author{V. H. Binh}
\email{vhbinh@iop.vast.vn}
\affiliation{\IOP} 

\author{H. T. Hung}
\email{hathanhhung@hpu2.edu.vn}
\affiliation{\SPtwo}

\author{Duong Van Loi}
\email{loi.duongvan@phenikaa-uni.edu.vn (corresponding author)}
\affiliation{\AdrHEPC}

\date{\today}

\begin{abstract}
We extend the $SU(3)_C \times SU(3)_L \times U(1)_X$ model with neutral leptons (331LHN) by introducing scalar leptoquarks. We determine the particle content of the leptoquark multiplets and their Yukawa interactions with fermions. We find that a singlet leptoquark can fully account for the $4.2\sigma$ discrepancy in the muon anomalous magnetic moment $\Delta a_\mu^{2021}$. The corresponding leptoquark mass is constrained to be $m_S \gtrsim 1.8$~TeV, consistent with current LHC bounds. We further consider the updated $\Delta a_\mu^{2025}$ based on recent lattice QCD results, which strengthen the lower bound to $m_S \gtrsim 6$~TeV. Combining $\Delta a_\mu$ with low-energy leptonic observables, including charged lepton flavor violation and the $\mu$--$e$ conversion rate, we constrain the viable parameter space. The allowed leptoquark Yukawa couplings exhibit a normal hierarchical pattern under all constraints. We also investigate the collider phenomenology of the singlet leptoquark, showing that its QCD-driven pair production leads to suppressed signal rates at the LHC for multi-TeV masses, while future hadron colliders can significantly extend the discovery reach.
\end{abstract}

\maketitle

\section{Introduction}
The Standard Model (SM) of particle physics remains the most successful and rigorously tested theoretical framework for describing the fundamental constituents of matter and their interactions. Nevertheless, several profound questions remain unanswered within the SM, such as the nature of dark matter, the origin of nonzero neutrino masses, and the replication of fermion families~\cite{ParticleDataGroup:2024cfk, McDonald:2016ixn, Kajita:2016cak, Planck:2018vyg}.

In recent years, a growing set of experimental anomalies spanning both high- and low-energy regimes has emerged, challenging the predictions of the SM. These include the long-standing discrepancy in the muon anomalous magnetic moment~\cite{Muong-2:2021ojo, AOYAMA20201, ALIBERTI20251}, tensions in precision measurements of the $W$ boson mass~\cite{CDF:2022hxs}, and persistent deviations in rare semileptonic $B$-meson decays, such as $R(K)$ and $R(K^*)$~\cite{LHCb:2015gmp,LHCb:2017smo,LHCb:2017rln,LHCb:2019hip,LHCb:2020lmf}. Additional hints of new physics include intriguing low-energy nuclear transitions that may point to a possible $\sim 17$~MeV boson (the so-called X17 anomaly)~\cite{Krasznahorkay:2015iga}. Moreover, the absence of clear signals in direct searches at the LHC has intensified interest in indirect or non-resonant probes, such as deviations in high-$p_T$ kinematic distributions~\cite{Buss:2022lxw,l36g-6f46}, which could reveal the presence of heavy, inaccessible particles through quantum effects.

These anomalies have stimulated intense theoretical and experimental efforts to uncover their underlying origins. Among the proposed explanations, models based on the $SU(3)_L \times U(1)_X$ extension of the SM stand out as a particularly compelling framework~\cite{Pisano1992, Frampton1992, Foot1993, Singer1980, Foot1994, Long1996,Diaz2005}. This class of models offers potential explanations for several outstanding issues, such as electric charge quantization, the number of fermion families, and the origin of neutrino masses~\cite{CarcamoHernandez2019}. In addition, these models exhibit a number of distinctive features, including the possibility of tree-level CP violation, the presence of flavor-changing neutral currents at tree level~\cite{Peccei1977,Promberger2007}, potential violations of lepton flavor universality~\cite{Lindner:2016bgg, Huong:2025qkc}, as well as nontrivial effects in electroweak precision observables such as the $W$-boson mass~\cite{VanLoi:2022eir}.

Leptoquarks (LQs), hypothetical scalar or vector bosons that carry both baryon and lepton numbers, have attracted significant attention over the past decades~\cite{Dorsner:2016wpm}. Originally introduced in the Pati--Salam model~\cite{PhysRevD.10.275}, their interactions are typically constrained by a generalized fermion number $F = 3B + L$, where $B$ and $L$ denote the baryon and lepton numbers, respectively~\cite{Pati:1975ca}. The conservation of $F$ allows for renormalizable and gauge-invariant couplings between quarks and leptons. The direct quark--lepton interactions mediated by LQs provide a distinctive portal to physics beyond the SM, leading to testable signatures at the LHC and future high-intensity experiments. It is therefore well motivated to explore the role of leptoquark interactions within the framework of $SU(3)_L$-based models.

The anomalous magnetic moment of the muon, $a_\mu = \left(\frac{g-2}{2}\right)_\mu$, has long been regarded as one of the most sensitive probes of physics beyond the SM and, until recently, one of its most persistent anomalies. For more than two decades, experimental measurements~\cite{Muong-2:2021ojo} of $a_\mu$ have consistently exhibited a $(4$--$5)\sigma$ deviation from the SM predictions~\cite{AOYAMA20201}:
\begin{align}
a_\mu^{\text{Exp},2021} &= 116592059(22)\times 10^{-11}, \hs 
a_\mu^{\text{SM}} = 116591802(2) \times 10^{-11}, \\
\Delta a_\mu^{2021} &= a_{\mu}^{\text{Exp},2021} - a_\mu^{\text{SM}} = 251(59)\times 10^{-11}.
\end{align}

However, a recent study~\cite{ALIBERTI20251} finds no significant discrepancy between the SM prediction and the experimental measurement of the muon anomalous magnetic moment $a_\mu$ at the current level of precision. This shift stems from major improvements in the evaluation of the hadronic light-by-light contribution, based on both data-driven dispersive approaches and lattice QCD calculations. As a result, the updated SM prediction is
\begin{equation}
	a_\mu^{\text{SM,\,update}} = 116592033(62)\times 10^{-11} \; (530~\text{ppb}).
\end{equation} 
Comparing this with the latest experimental average, obtained by combining the E821 experiment and the Fermilab E989 Run~1--6 results, the discrepancy becomes
\be
	\Delta a_\mu^{2025} = a_\mu^{\text{Exp},2025} - a_\mu^{\text{SM,\,update}} = (38 \pm 63) \times 10^{-11}.
\ee
This does not imply that new physics is ruled out. Rather, it underscores the crucial role of theoretical uncertainties. In particular, independent confirmation remains essential given the significant QCD-related uncertainties, especially in the hadronic vacuum polarization and hadronic light-by-light contributions. Moreover, the Fermilab Muon $g$-2 experiment has not yet reached its design precision of $127~\text{ppb}$ ($\simeq 16 \times 10^{-11}$). Therefore, in this work, we adopt a cautious approach by considering both the $\Delta a_\mu^{2021}$ result and the most recent updates.

Our work is organized as follows. In Section~\ref{sec2}, we introduce the theoretical framework of the model. In Section~\ref{sec3}, we define the leptoquark sector, specifying its representation content and the relevant Yukawa couplings to SM fermions. In Section~\ref{sec4}, we compute the contributions of the singlet leptoquark to $a_\mu$ and explore the parameter space in light of both the 2021 and updated 2025 experimental results and SM predictions. In Section~\ref{sec5}, we evaluate the contributions of the singlet leptoquark to charged lepton flavor-violating (CLFV) processes, including the radiative decays $l_i \to l_j \gamma$ and $\mu$--$e$ conversion in nuclei, and constrain the model parameter space by combining bounds from $a_\mu$ and CLFV observables. In Section~\ref{sec6}, we study the collider phenomenology of the singlet leptoquark at proton--proton colliders, focusing on QCD pair production and the resulting signatures. Finally, Section~\ref{sec7} summarizes our findings and discusses the implications for future tests.

\section{\label{sec2}Particle content of the model}
In the $SU(3)_C \times SU(3)_L \times U(1)_X$ (331LHN) model, a neutral lepton is introduced to form the third component of the leptonic triplets~\cite{PhysRevD.47.2918,Cogollo_2014},
\begin{equation}
	L_{aL}= \begin{pmatrix}
		\nu_{aL}     \\
		e_{aL}       \\
		N_{aL} \\ 
	\end{pmatrix} \sim (1,3,-1/3), \quad e_{aR}\sim (1,1,-1), \quad N_{aR}\sim (1,1,0),
\end{equation}
where $a = 1,2,3$ labels the three lepton generations.

Anomaly cancellation requires that the number of triplets equals the number of antitriplets. Due to the color multiplicity of quarks, the quark sector must contain two antitriplets and one triplet. Therefore, the first two quark families are assigned to transform as antitriplets, while the third family transforms as a triplet. We then have:
\begin{align}
	Q_{iL} &= \left (
	\begin{array}{c}
		d_{iL} \\
		-u_{iL} \\
		d^{\prime}_{iL}
	\end{array}
	\right )\sim(3\,,\,3^*\,,\,0),\hs Q_{3L} = \left (
	\begin{array}{c}
		u_{3L} \\
		d_{3L} \\
		u^{\prime}_{3L}
	\end{array}
	\right )\sim(3\,,\,3\,,\,1/3),\\
	u_{iR}&\sim(3,1,2/3),\hs d_{iR}\,\sim(3,1,-1/3),\hs d^{\prime}_{iR}\,\sim(3,1,-1/3), \\
	u_{3R}&\sim(3,1,2/3),\hs d_{3R}\,\sim(3,1,-1/3),\hs u^{\prime}_{3R}\,\sim(3,1,2/3),
	\label{quarks} 
\end{align}
where $i = 1,2$ labels the first two quark generations.

The original scalar sector of the model consists of three scalar triplets,
\begin{eqnarray}
	\eta = \left (
	\begin{array}{c}
		\eta^0 \\
		\eta^- \\
		\eta^{\prime 0}
	\end{array}
	\right ),\hs\rho = \left (
	\begin{array}{c}
		\rho^+ \\
		\rho^0 \\
		\rho^{\prime +}
	\end{array}
	\right ),\hs
	\chi = \left (
	\begin{array}{c}
		\chi^0 \\
		\chi^{-} \\
		\chi^{\prime 0}
	\end{array}
	\right ),
	\label{scalarcont} 
\end{eqnarray}
where $\eta$ and $\chi$ transform as $(1,3,-1/3)$, while $\rho$ transforms as $(1,3,2/3)$. After the spontaneous breaking of the $SU(3)_L \times U(1)_X$ symmetry, the neutral scalar fields develop vacuum expectation values (VEVs): $\langle \eta^0 \rangle = v_\eta$, $\langle \rho^0 \rangle = v_\rho$, and $\langle \chi^{\prime 0} \rangle = v_{\chi'}$.

Imposing a $Z_2$ discrete symmetry, the relevant Yukawa interactions in the lepton sector are given by~\cite{Mizukoshi:2010ky}
\begin{equation}
	{\cal L} \supset g_{ab}\bar f_{aL} \rho e_{bR}+g^{\prime}_{ab}\bar{f}_{aL}\chi N_{bR}+ \text{h.c.}, 
	\label{yukawa1}
\end{equation}
\begin{equation}
	{\cal L} \supset \frac{y_{ab}}{\Lambda}\bar{f^c}_{aL}\eta^*\eta^{\dagger} f_{bL} + \text{h.c.},
	\label{numasses}
\end{equation}
where $\rho$, $\eta$, and $\chi$ denote the scalar triplets introduced in Eq.~(\ref{scalarcont}). The charged leptons acquire masses from the first term in Eq.~(\ref{yukawa1}), whereas neutrino masses are generated via the dimension-five effective operator in Eq.~(\ref{numasses}). Throughout this work, the leptonic Yukawa matrices are taken to be diagonal and follow a normal hierarchy. In the quark sector, the Yukawa interactions are given by
\begin{eqnarray}
	\cal L  &\supset& \alpha_{ij} \bar Q_{iL}\chi^* d^{\prime}_{jR} +\al_{33} \bar Q_{3L}\chi u^{\prime}_{3R} + g_{ia}\bar Q_{iL}\eta^* d_{aR} \nonumber \\
	&&+h_{3a} \bar Q_{3L}\eta u_{aR} +g_{3a}\bar Q_{3L}\rho d_{aR}+h_{ia}\bar Q_{iL}\rho^* u_{aR} + \text{h.c.}, \label{yukawa2}
\end{eqnarray}
where $i,j = 1,2$ and $a = 1,2,3$. The SM quark masses are recovered in the usual way once $v_{\rho} = v_{\eta} = v$, with $v = v_{\text{SM}}/\sqrt{2}$. The masses of the exotic quarks $d^{\prime}_i$ and $u^{\prime}_3$ originate from the first two terms in Eq.~(\ref{yukawa2}) and are given by
\be
M_{d^{\prime}_i}=\frac{\alpha_{ii}}{\sqrt{2}}v_{\chi^{\prime}}, \hs M_{u^{\prime}_3}=\frac{\alpha_{33}}{\sqrt{2}}v_{\chi^{\prime}}.
\label{mquarks}
\ee

We now consider the gauge boson sector. The model contains nine electroweak gauge bosons: eight $W_\mu^i$ $(i=1,\dots,8)$ of $SU(3)_L$ and one $B_\mu$ of $U(1)_X$. These gauge bosons arise from the kinetic terms of the scalar fields,
\be 
\mathcal{L}_{\text{Higgs}} = \sum_{H=\eta,\rho,\chi} (D^\mu H)^\dagger D_\mu H \,.
\label{kineticGB}
\ee
The mass eigenstates of the charged gauge bosons are given by
\bea 
W^{\pm}_{\mu } &=& \frac{1}{\sqrt{2}}\left( W^1_{\mu}\mp i W^2_{\mu}\right), \hs
W^{\prime \pm}_{\mu} = \frac{1}{\sqrt{2}}\left( W^6_{\mu}\pm i W^7_{\mu}\right), \\
U^{0}_{\mu} &=& \frac{1}{\sqrt{2}}\left( W^4_{\mu}- i W^5_{\mu}\right), \hs
U^{0\dagger}_{\mu}=\frac{1}{\sqrt{2}}\left( W^4_{\mu}+ i W^5_{\mu}\right),
\label{j234}
\eea
with the corresponding masses
\be 
m^2_{W} = \frac{g^2}{4}(v_\eta^2+v_\rho^2), \hs 
m^2_{U^0}=\frac{g^2}{4}(v_{\chi^\prime}^2+v_\eta^2), \hs 
m^2_{W^{\prime \pm}}=\frac{g^2}{4}(v_{\chi^\prime}^2+v_\rho^2)\,,
\label{mWXY} 
\ee
where $g$ is the $SU(3)_L$ gauge coupling.

About the mass mixing matrix of neutral gauge bosons, this matrix is diagonalized by an orthogonal transformation (see Appendix A of Ref.~\cite{Binh:2020aal} for details). The physical neutral gauge bosons consist of the photon $A$ and two massive states $Z$ and $Z^\prime$, whose masses are given by
\begin{align}
m^2_A &= 0,\\
m^2_Z &= \frac{g^2(9g^2 + 2 g_X^2)(v_\eta^2 v_\rho^2 + v_\rho^2 v_{\chi^\prime}^2 + v_{\chi^\prime}^2 v_\eta^2)}{2\left[18g^2(v_\eta^2 + v_\rho^2 + v_{\chi^\prime}^2)+g_X^2 (v_\eta^2 + 4v_\rho^2 + v_{\chi^\prime}^2)\right]}, \\
m^2_{Z^\prime}
&= \frac{g^2}{3} (v_\eta^2 + v_\rho^2 + v_{\chi^\prime}^2) + \frac{g_X^2}{54}(v_\eta^2 + 4v_\rho^2 + v_{\chi^\prime}^2) - m^2_Z,\label{mAZZp}
\end{align}
where $g_X$ denotes the gauge coupling of the $U(1)_X$ group. In the limit $v_{\chi^\prime} \gg v_\eta, v_\rho$, the $Z^\prime$ mass is dominated by the symmetry-breaking scale $v_{\chi^\prime}$.

Among these nine gauge bosons, four correspond to the SM states ($W^{\pm}$, $Z$, and $\gamma$), while the remaining five ($W^{\prime \pm}$, $U^0$, $U^{0\dagger}$, and $Z^\prime$) are new states predicted by the 3-3-1 extension. The SM gauge bosons reproduce the usual phenomenology, whereas the new gauge bosons acquire masses as given in Eqs.~(\ref{mWXY}) and (\ref{mAZZp}).

\section{\label{sec3}Leptoquark: representation content and Yukawa interactions}
The most general Lagrangian for a scalar leptoquark $S$, classified according to the fermion number $F = 3B + L$ with $F = 0$ or $|F| = 2$, can be written as
\begin{align}
	\mathcal{L}^{F=0}&=\bar{q}_i \left(y^{ij}_R P_R  + y_L^{ij} P_L\right)l_j S + \text{h.c.}, \\
	\mathcal{L}^{|F|=2}&=\overline{q_i^{c}} \left(\lambda^{ij}_R P_R  + \lambda_L^{ij} P_L\right) l_jS + \text{h.c.},
\end{align}
where $q_i^c = C\bar{q}_i^{T}$ denotes the charge-conjugated quark field.

From the quark and lepton content of the model, one can construct the following fermion bilinears:
\begin{align}
	\bar{Q}^c_{iL} L_{aL} & \sim  (3, 3^*, 0) \times (1\,,\,3,-1/3)  \sim (3\,,\,1 \oplus 8\,,\,-1/3), \\
	\bar{Q}_{3L}^c L_{aL} & \sim (3, 3, 1/3) \times (1 , 3, -1/3)   \sim (3, 3^* \oplus 6 , 0), \\
	\bar{u}_{aR} L_{bL} &\sim  (3^*, 1, -2/3)  \times (1 ,3, -1/3) \sim (3^* ,3, -1), \\	
	\bar{d}_{aR} L_{bL}  &\sim (3^* , 1, 1/3)  \times (1, 3, -1/3) \sim (3^*, 3, 0), \\
	\bar{u}^c_{aR} e_{bR}  &\sim (3, 1, 2/3) \times (1, 1, -1)  \sim (3 , 1, -1/3),	 \\
	\bar{d}^c_{aR} e_{bR} & \sim  (3, 1, -1/3) \times   (1, 1, -1) \sim (3,  1,-4/3),
	\label{LQbilinearterms}
\end{align}
where we have used the tensor product decomposition $3 \otimes 3 = 3^* \oplus 6$ and $3^* \otimes 3 = 1 \oplus 8$. To construct gauge-invariant Yukawa interactions, these bilinears require the following scalar leptoquark representations:
\begin{align}
	S_8 &\sim (3, 8, -1/3),\hs S_6 \sim (3,6,0),\\ 
	S_3 &\sim(3^*, 3, 0),\hs \tilde{S_3} \sim (3^* ,3, -1), \\
	S &\sim (3 ,1 , -1/3),\hs \tilde{S} \sim (3 ,1, 4/3).
	\label{LQrep}
\end{align}

With the above leptoquark representations, the Yukawa Lagrangian can be written as
\begin{align}
	{\mathcal{L}} &\supset  Y_{3a} \bar{Q}_{3L}^c L_{aL} S_6^\dagger   + Y_{ia} \bar{Q}^c_{iL} L_{aL} S_8^\dagger   + \text{h.c.}, \\
	{\mathcal{L}}&\supset h_{3a} \epsilon_{ijk} ( \bar{Q}_{3L}^c)_i (L_{aL})_j  (S_3)_k + \tilde h_{ab} \bar{d}_{aR}  L_{bL} \tilde S_3   + \text{h.c.}, \\
	{\cal L} &\supset \tilde \lambda_{ia}\bar Q^c_{iL} L_{aL} S^\dagger   + y_{ab}\bar u^c_{aR} e_{bR} S^\dagger + \text{h.c.},
	\label{LQ331LHN}
\end{align}
where $i = 1,2$ and $a,b = 1,2,3$.

The components of the sextet $S_6$ and the octet $S_8$ are given by
\begin{eqnarray}
	S_6=\begin{pmatrix}
		\tilde{\sigma_1}^{2/3} & \sigma_2^{-1/3} & \sigma_3^{2/3} \\
		\sigma_2^{-1/3} & \tilde{\sigma_2}^{-4/3}  & \sigma_4^{-1/3} \\
		\sigma_3^{2/3}& \sigma_4^{-1/3} &  \tilde \sigma_3^{2/3}  
			\label{Sextet} 
			\end{pmatrix},
		 \hs 
		 S_8=\begin{pmatrix}
	\tilde \omega_1^{1/3} & \omega_2^{-1/3} & \omega_3^{2/3} \\
	\omega_4^{-1/3} &\,\tilde \omega_2^{-5/3}  & \omega_4^{-1/3} \\
	\omega_5^{2/3}& \omega_6^{-1/3} &  \, \tilde \omega_3^{1/3}  
		 \end{pmatrix}.
	\label{Octet} 
\end{eqnarray}
The triplet leptoquarks $S_3$ and $\tilde{S}_3$ are given by
\begin{eqnarray}
	S_3 = \left (
	\begin{array}{c}
		S_3^{1/3} \\
		S_3^{-2/3} \\
		S_3^{1/3}
	\end{array}
	\right ),
	\hs\tilde{S_3} = \left (
	\begin{array}{c}
		\tilde S_3^{4/3} \\
		\tilde S_3^{\prime 1/3} \\
		\tilde S_3^{4/3}
	\end{array}
	\right )
	\label{tripletLQ} .
\end{eqnarray}

The Yukawa Lagrangian involving the singlet leptoquark $S$ is given by
\bea
	{\cal L} &\supset & \tilde y_L^{ia}\bar Q^c_{iL} L_{aL} S^\dagger   + y_R^{ab}\bar u^c_{aR} e_{bR} S^\dagger + \text{h.c.} \crn
	&=& \tilde y_L^{ia}(\bar d^c_{iL}  \nu_{aL}-\bar u^c_{iL} e_{aL}+ \bar d'^c_{iL} N_{aL} ) S^\dagger + y_R^{ab}\bar u^c_{aR} e_{bR} S^\dagger + \text{h.c.}
	\label{YukawaLS} 
\eea
The singlet scalar leptoquark $S$ induces chirality-flipping interactions, leading to the dominant contribution to the anomalous magnetic moment $(g-2)_\mu$ \cite{Stockinger:2022ata,YU2022115674}. For this reason, our analysis focuses on the physical properties and phenomenological implications of $S$.

The physical (mass eigenstate) quark fields $\hat{u}_{a}=(u\,\,c\,\,t)$ and $\hat{d}_{a}=(d\,\,s\,\,b)$ are related to the gauge eigenstates via the unitary transformations $\hat{u}_{L}=V^{u\dagger}_{L} u_{L}$ and $\hat{d}_{L}=V^{d\dagger}_{L} d_{L}$, with the CKM matrix given by $V_{\text{CKM}}=V^{u}_{L} V^{d\dagger}_{L}$. Without loss of generality, we work in a basis where the right-handed charged lepton sector is diagonal. In this basis, the Yukawa interactions between the singlet leptoquark $S$ and the fermion mass eigenstates are given by
\begin{equation}
	{\cal L}=\bar{\hat{u}}^c_{a}(y_L^{ab}P_L + y_R^{ab}P_R)e_b S + y^{\prime ab }_L \bar{\hat{d}}^c_{aL} \nu_{aL}S +  \tilde{y}^{ia} \bar \nu_{aR} d^{\prime}_{iL}S +  \text{h.c.},
	\label{QS-Y}
\end{equation}
where the effective couplings are related to the gauge-basis parameters by $y_L^{ab}=-\tilde{y}_L^{ai}(V^u_L)_i^b$ and $y_L^{\prime ab}=\tilde{y}_L^{ ai}(V^d_L)_{i}^b$. 

\section{\label{sec4}Contributions of the singlet leptoquark $S$ to $\Delta a_\mu$}
\begin{figure}[t]
	\centering
	\includegraphics[width=0.9\columnwidth]{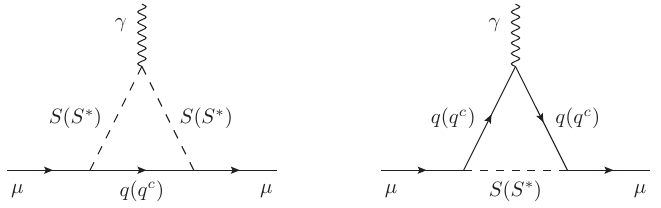}
	\caption{One-loop diagrams contributing to $\Delta a_\mu$ induced by the singlet leptoquark $S$.}
	\label{Scont}
\end{figure}

Since the muon anomalous magnetic moment corresponds to a chirality-flipping operator,
\begin{align}
	 \bar{\mu}_R \sigma_{\mu \nu} \mu_L + \text{h.c.},
\end{align} 
any interaction that induces such a chirality flip can generate a sizable contribution to $a_\mu$ \cite{Bigaran:2020jil,PhysRevD.95.055018,Crivellin:2022wzw}. Accordingly, we focus on the interactions between the singlet leptoquark $S$ and charged fermions that induce a chirality flip of the muon.

The one-loop contributions to $\Delta a_\mu$ induced by the singlet leptoquark $S$ are given by \cite{Dorsner:2016wpm, Dorsner:2019itg, Bigaran:2020jil, He:2021yck,PhysRevD.108.095027}:
\begin{equation}
	\Delta a_\mu^{}=\delta a_\mu^{QED}(m_\mu) \frac{m^2_\mu}{48 \pi^2 m^2_S}\left( \frac{m_q}{m_\mu}y_L^{a2} y_R^{a2}L_1(x_q)+\frac{ (y_L^{a2})^2 + (y_R^{a2})^2}{4}L_2(x_q) \right), 
	\label{Damu331LHN}
\end{equation}
where $x_q = m_q^2/m_S^2$, and 
\begin{align}
	\delta a_\mu^{QED}(m_\mu) & = 1 + \frac{4\alpha}{\pi}\ln\left(\frac{m_\mu}{m_{S}}\right),\\
	L_1(x)&=4F_F(x)-F_C(x), \hs L_2(x)=2F_E(x)-F_B(x),\\
	F_F(x)&=\frac{3(-3+4x-x^2-2\ln x)}{2(1-x)^3}, \\
	F_C(x)&=\frac{3(1-x^2+2x\ln x)}{(1-x)^3}\\
	F_E(x)&=\frac{2(2+3x-6x^2+x^3+6x\ln x)}{(1-x)^4}, \\
	F_B(x)&=\frac{2(1-6x+3x^2+2x^3-6x^2 \ln x)}{(1-x)^4}.
\end{align}
The expression in Eq.~(\ref{Damu331LHN}) exhibits a clear chirality enhancement, with $\Delta a_\mu$ proportional to the quark mass $m_q$. As a result, the dominant contribution arises from the top quark. In the following, we therefore focus on the leading contribution corresponding to $q = t$ (i.e., $a = 3$ in Eq.~(\ref{Damu331LHN})). $\Delta a_\mu$ can be approximated as
\begin{equation}
	\Delta a_\mu^{}=\delta a_\mu^{QED}(m_\mu) \frac{m^2_\mu}{48 \pi^2 m^2_S}\left( \frac{m_t}{m_\mu}y_L^{a2} y_R^{a2}L_1(x_q) \right), 
	\label{Damu331LHNapprox}
\end{equation}
where $x_t = m_t^2/m_S^2$.

\begin{figure}
	\includegraphics[width=0.415\linewidth]{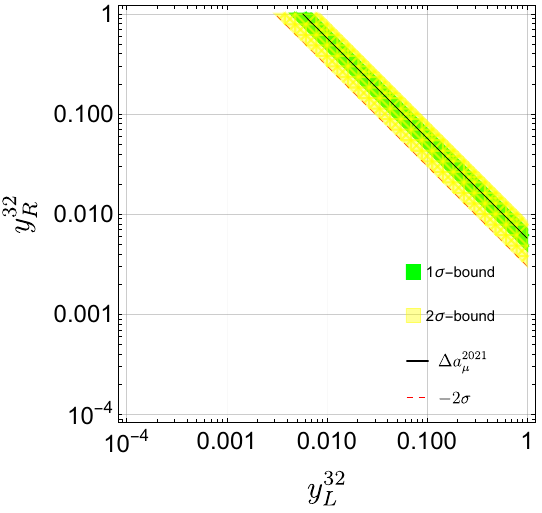}
	\includegraphics[width=0.55\linewidth]{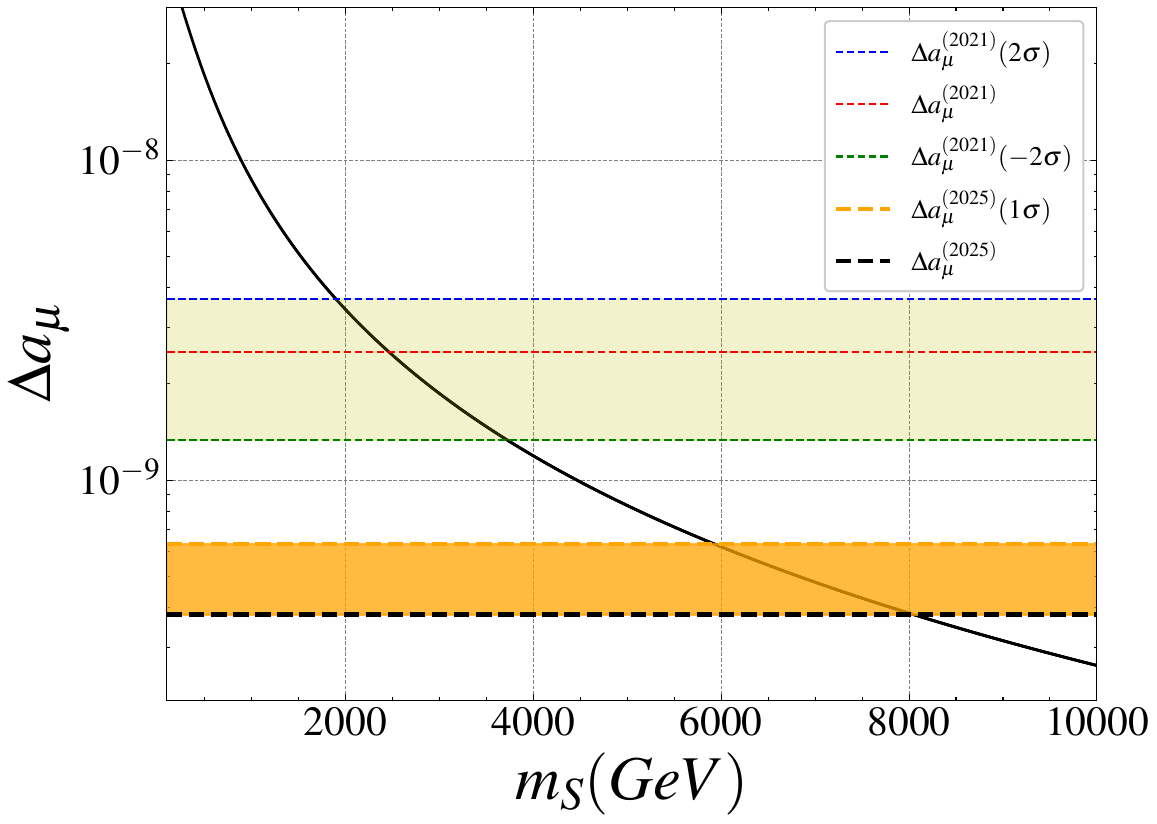}
\caption{Allowed parameter space for the one-loop contributions to $\Delta a_\mu^{2021}$ and $\Delta a_\mu^{2025}$ induced by the singlet leptoquark $S$.}
	\label{Damu}
\end{figure}

The $2\sigma$ band of $\Delta a_\mu^{2021}$ corresponds to the condition
\begin{equation}
	3.3\times 10^{-3} < y_L^{32} y_R^{32} < 9.3 \times 10^{-3},
		\label{Damu2021cond}
\end{equation}
while the $1\sigma$ band of $\Delta a_\mu^{2025}$ leads to
\begin{equation}
	y_L^{32} y_R^{32} < 2.54 \times 10^{-3}.
	\label{Damu2025cond}
\end{equation}
In the left plane of Fig.~\ref{Damu}, we present $\Delta a_\mu^{2021}$ (using the 2021 experimental value) on a logarithmic scale, highlighting the allowed region in the $(y_L^{32}, y_R^{32})$ parameter space. The figure displays the regions compatible with the experimental value of $\Delta a_\mu^{2021}$ at the $1\sigma$ (green) and $2\sigma$ (yellow) confidence levels.

In the right plane of Fig.~\ref{Damu}, we show the dependence of $\Delta a_\mu^{2021,2025}$ on the singlet leptoquark mass $m_S$, imposing the constraints given in Eqs.~(\ref{Damu2021cond}) and (\ref{Damu2025cond}). As illustrated in the right plane of Fig.~\ref{Damu}, the 2021 data yield an allowed band requiring $m_S \gtrsim 1.8~\mathrm{TeV}$, while the projected 2025 sensitivity (shown as the $1\sigma$ orange band) pushes this lower bound to $m_S \gtrsim 6~\mathrm{TeV}$. This lower bound arises from the scaling $\Delta a_\mu \propto 1/m_S^2$, implying that smaller contributions to $\Delta a_\mu$ allow for heavier leptoquarks.

\section{\label{sec5}Charged lepton flavor violation induced by the singlet leptoquark $S$}
The branching ratio for the lepton flavor-violating radiative decay $l_i \rightarrow l_j \gamma$, similarly to the case of the muon anomalous magnetic moment, is dominated by chirality-enhanced contributions. The branching ratio for the process $l_i \rightarrow l_j \gamma$ is given by \cite{He:2021yck,PhysRevD.108.095027}
\begin{align}
	\mathrm{Br}(l_i \rightarrow l_j \gamma) &= \frac{m^5_{l_i}}{16\pi\Gamma_{l_i}}\left(|A^L_2|^2 + |A^R_2|^2\right),
	\label{BRliljg}
\end{align}
where $\Gamma_\mu = 2.996 \times 10^{-19}~\mathrm{GeV}$ and $\Gamma_\tau = 2.267 \times 10^{-12}~\mathrm{GeV}$ \cite{PhysRevD.110.030001}. The dipole form factors $A_2^{L,R}$ are given by
\begin{align}
	A^L_2 &= -\frac{1}{16\pi^2}\frac{e}{6m^2_S}\left(\frac{m_q}{m_{l_i}}y_L^{qi}y_R^{qj}L_1(x_q) + \frac{1}{4} y_R^{qi}y_R^{qj} L_2(x_q)\right),  \\
	A^R_2 &= -\frac{1}{16\pi^2}\frac{e}{6m^2_S}\left(\frac{m_q}{m_{l_i}} y_R^{qi}y_L^{qj}L_1(x_q) + \frac{1}{4} y_L^{qi}y_L^{qj}L_2(x_q)\right),
	\label{A2LR}
\end{align}
with $x_q = m_q^2/m_S^2$.

Considering only the top-quark contribution in Eq.~(\ref{BRliljg}), we obtain the following constraint:
\begin{equation} 
	|y_R^{3i}y_L^{3j}|^2 + |y_L^{3i}y_R^{3j}|^2 \leq \frac{ \mathrm{Br}(l_i \to l_j \gamma) \Ga_{l_i}   m_S^4}{1.268 m_i^3(1-0.167 \ln{m_S})^2}.
	\label{yLRcond}
\end{equation}
In the following, we will use the constraint in Eq.~(\ref{yLRcond}), together with the experimental bounds from the muon anomalous magnetic moment $\Delta a_\mu$ and the branching ratios of the decays $l_i \to l_j \gamma$, to determine the phenomenologically allowed parameter space for the leptoquark couplings.

\subsection{$\mu \to e \gamma$}
We now consider the impact of CLFV, in particular the $\mu \to e \gamma$ process, on the leptoquark couplings. This process involves the muon and is governed by a dipole interaction, dominated by chirality-flipping contributions. We analyze two distinct coupling scenarios:
\begin{itemize}
\item[-] Symmetric coupling: the left- and right-handed leptoquark couplings satisfy $y_L^{32} = y_R^{32}$.
\item[-] Asymmetric coupling: $y_L^{32} \neq y_R^{32}$, with the ratio defined as $r^{3i} = y_L^{3i}/y_R^{3i}$.
\end{itemize}

\begin{figure}[ht!]
	\includegraphics[width=0.5\linewidth]{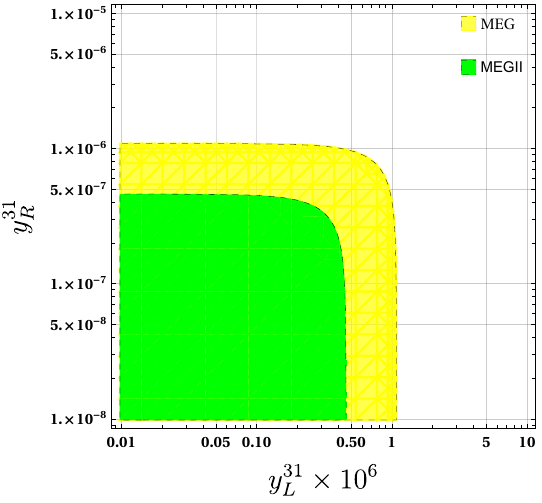}
	\includegraphics[width=0.48\linewidth]{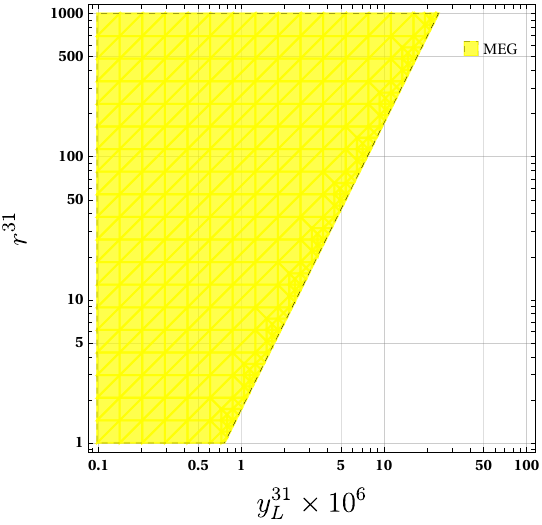}
\caption{Allowed parameter space consistent with the experimental bound on $\mu \to e \gamma$.}
	\label{muegFig}
\end{figure}

In the left panel of Fig.~\ref{muegFig}, we analyze the branching ratio of $\mu \to e \gamma$ in the case of symmetric leptoquark couplings, taking into account both current and projected experimental constraints. The MEG experiment \cite{MEG:2020zxk} sets an upper bound $\mathrm{Br}(\mu \to e \gamma) \leq 4.2 \times 10^{-13}$, while the upgraded MEG II experiment \cite{MEGII:2018kmf} is expected to improve the sensitivity down to $6.0 \times 10^{-14}$. Using the constraint from $\Delta a_\mu$ (i.e., $y_L^{32} y_R^{32} \leq 2.54 \times 10^{-3}$ from Eq.~(\ref{Damu2025cond})), the leptoquark Yukawa couplings $y_L^{31}$ and $y_R^{31}$ are required to be at most of order $\mathcal{O}(10^{-6})$ in order to satisfy these bounds.

To quantify the hierarchy of leptoquark couplings, we define $h^{21} = y_L^{32}/y_L^{31}$. Combining the $\Delta a_\mu$ constraint with the assumption of symmetric left- and right-handed couplings, we obtain $h^{21} \simeq 10^4$. Taking the lower bound $y_L^{32} \simeq 5 \times 10^{-2}$ from $\Delta a_\mu$, this implies $y_L^{31} \simeq 5 \times 10^{-6}$.

In the right panel of Fig.~\ref{muegFig}, we present the MEG constraint on $\mathrm{Br}(\mu \to e \gamma)$ in the $(y_L^{31}, r^{31})$ plane, where $r^{31} = y_L^{31}/y_R^{31}$. In the symmetric limit $r^{31} \to 1$, we recover $y_L^{31} \simeq y_R^{31} \simeq 10^{-6}$. As $r^{31}$ increases, the allowed value of $y_L^{31}$ also increases; for instance, for $r^{31} = 10^3$, the upper bound reaches $y_L^{31}\simeq 2\times 10^{-5}$.

\subsection{$\tau \to \mu \gamma$}
\begin{figure}[ht!]
	\includegraphics[width=0.49\linewidth]{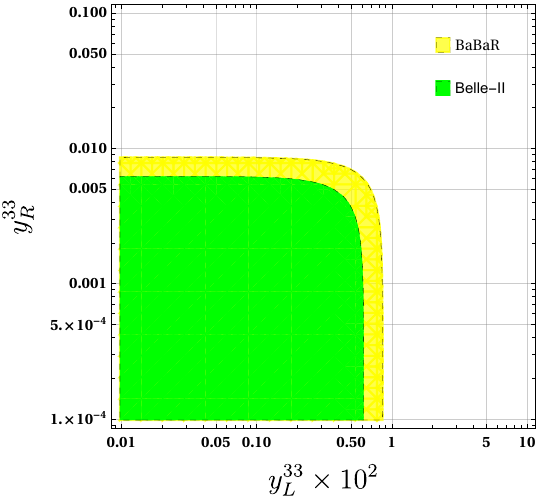}
	\includegraphics[width=0.47\linewidth]{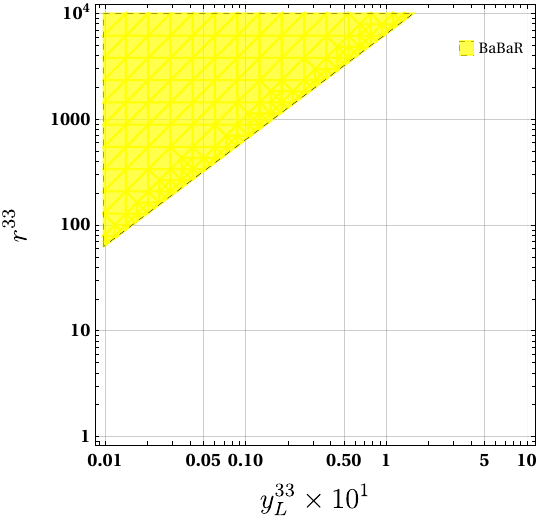}
\caption{Allowed parameter space consistent with the experimental bounds on $\tau \to \mu \gamma$.}
	\label{taumugFig}
\end{figure}

Next we examined  the allowed parameter space for  $\tau \to \mu \gamma$ decay as illustrated in Fig.~\ref{taumugFig}. Like the previous $\mu \to e \gamma $ decay process, this decay is also dipole induced and involve muon therefore get constraint from muon $\Delta a_\mu$.

We analyze the impact of the singlet leptoquark on the couplings in two scenarios: symmetric and asymmetric. The current bound from the BaBar experiment \cite{PhysRevLett.104.021802} is $\mathrm{Br}(\tau \to \mu \gamma) \leq 4.4 \times 10^{-8}$, while the future sensitivity at Belle II \cite{Banerjee:2022xuw} is expected to reach $\mathrm{Br}(\tau \to \mu \gamma) \leq 6.9 \times 10^{-9}$. In the symmetric case (in the left plane of Fig.~\ref{taumugFig}), the couplings $y_L^{33}$ and $y_R^{33}$ are required to be at most of order $\mathcal{O}(10^{-2})$ in order to satisfy these bounds, given the constraints from $\Delta a_\mu$ in Eq.~(\ref{Damu2025cond}). In the asymmetric case, we define, without loss of generality, the ratio $r^{33} = y_L^{33}/y_R^{33} \geq 1$. The right plane of Fig.~\ref{taumugFig} shows the allowed values of $y_L^{33}$ and $r^{33}$, where we impose the BaBar bound on $\mathrm{Br}(\tau \to \mu \gamma)$ together with the constraints from $\Delta a_\mu$ on $y_{L,R}^{32}$ (Eq.~(\ref{Damu2025cond})) and the condition in Eq.~(\ref{yLRcond}). We find that $y_L^{33}$ can be as large as $\mathcal{O}(0.1)$. However, to satisfy the constraint in Eq.~(\ref{yLRcond}), the right-handed coupling must be significantly suppressed, $y_R^{33} = y_L^{33}/r^{33} \simeq 10^{-5}$.

\subsection{$\tau \to e \gamma$}
The condition in Eq.~(\ref{yLRcond}) can be rewritten as
\begin{equation} 
	y_L^{3i} y_R^{3i} y_L^{3j} y_R^{3j} \leq \frac{ \mathrm{Br}(l_i \to l_j \gamma) \Ga_{l_i}   m_S^4}{1.268 m_i^3(1-0.167 \ln{m_S})^2}
	\left[  \frac{r^{3i} r^{3j}}{(r^{3i})^2 + (r^{3j})^2}   \right],
	\label{4yLRcond}
\end{equation}
where $r^{3i}=y_L^{3i}/y_R^{3j}$.

\begin{figure}[ht!]
	\includegraphics[width=0.485\linewidth]{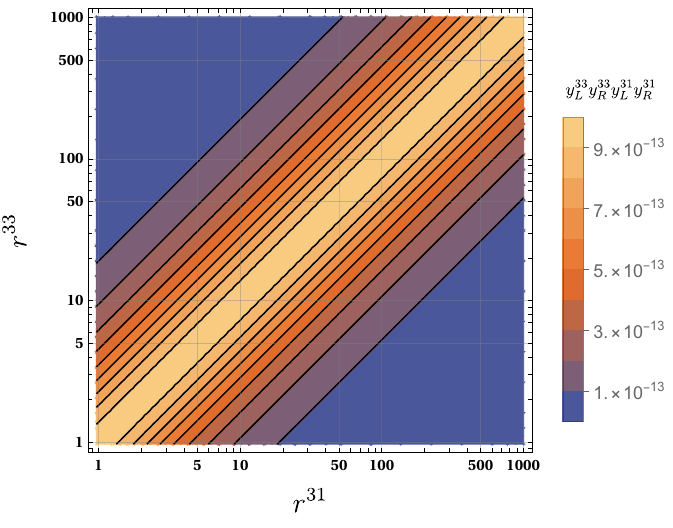}
	\includegraphics[width=0.495\linewidth]{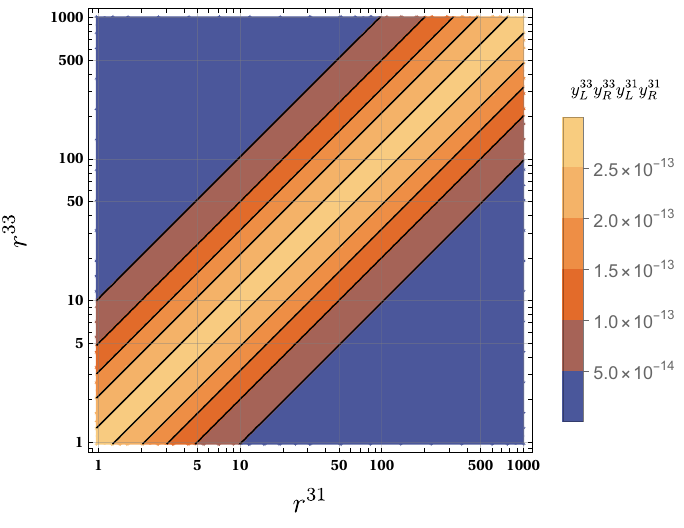}
\caption{Allowed parameter space consistent with the experimental bounds on $\tau \to e \gamma$.}
	\label{tauegFig}
\end{figure}

In Fig.~\ref{tauegFig}, we analyze the branching ratio $\mathrm{Br}(\tau \to e \gamma)$ under current and projected experimental sensitivities. The BaBar collaboration sets the strongest current limit, $\mathrm{Br}(\tau \to e \gamma) \leq 3.3 \times 10^{-8}$ \cite{PhysRevLett.104.021802}, while Belle II is expected to improve this sensitivity to $\mathrm{Br}(\tau \to e \gamma) \leq 9.0 \times 10^{-9}$ \cite{Banerjee:2022xuw}. Our analysis focuses on the dominant leptoquark contribution, proportional to the quartic product $y_L^{33} y_R^{33} y_L^{31} y_R^{31}$, and explores its dependence on the chirality asymmetry parameters $r^{31} = y_L^{31}/y_R^{31}$ and $r^{33} = y_L^{33}/y_R^{33}$.

The resulting allowed regions in the $(r^{31}, r^{33})$ plane are shown in Fig.~\ref{tauegFig}, with the BaBar constraint displayed in the left panel and the projected Belle II sensitivity in the right panel. The allowed parameter space remains consistent with other constraints, including Eq.~(\ref{yLRcond}), $\Delta a_\mu$, and the bounds from $\mu \to e \gamma$ and $\tau \to \mu \gamma$.

\subsection{$\mu-e$ conversion in nuclei}
\begin{figure}[ht!]
	\centering
	\includegraphics[width=0.4\columnwidth]{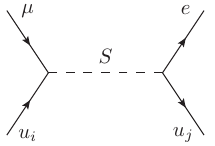}
	\caption{Tree-level diagram contributing to the $\mu$--$e$ conversion by the singlet leptoquark.}
	\label{mu2ediagram}
\end{figure}

The $\mu$--$e$ conversion process is of particular interest because, unlike radiative decays such as $\mu \to e \gamma$, it proceeds at tree level via leptoquark exchange, as illustrated in Fig.~\ref{mu2ediagram}. This leads to a significantly enhanced sensitivity to certain leptoquark couplings, especially those involving both quark and lepton flavors (e.g., $y_{L,R}^{21}$), and avoids the loop suppression that governs most CLFV processes. As a result, current and future experiments, such as SINDRUM II \cite{SINDRUMII:2006dvw}, Mu2e \cite{MISCETTI2025170257}, and COMET \cite{Fujii_2023}, place some of the most stringent constraints on leptoquark models compatible with the observed muon anomaly and other flavor observables.

The $\mu$--$e$ conversion rate is defined as~\cite{PhysRevD.66.096002}
\begin{align}
\mathrm{Cr}(\mu \text{--} e)= \frac{\big(\alpha_s y_R^{12} - \alpha_v y_L^{12}\big)^2 \big(y_L^{11}\big)^2 + [L \leftrightarrow R]}{4 m_S^{4}\Gamma_{\mathrm{capture}}}.
\label{eq:mec-overall}
\end{align}
For gold (Au) and aluminum (Al) targets, the relevant coefficients are given by
\begin{align}
\alpha_s &= 1.537\, m_{\mu}^{5/2} \quad (0.430\,m_{\mu}^{5/2}), \\
\alpha_v &= 0.280 \,m_{\mu}^{5/2} \quad (0.049\,m_{\mu}^{5/2}), \\
\Gamma_{\mathrm{capture}}
&= 8.849 \times 10^{-18}~\mathrm{GeV} \quad (0.464\times 10^{-18}~\mathrm{GeV}),\label{formfactors}
\end{align}
where the values in parentheses correspond to the aluminum target. The current limit from the SINDRUM II experiment for $\mu$--$e$ conversion in muonic gold is $\mathrm{Cr}(\mu\text{--}e,\mathrm{Au}) \leq 7.0 \times 10^{-13}$ \cite{SINDRUMII:2006dvw}. Future experiments such as Mu2e and COMET are expected to significantly improve the sensitivity for aluminum targets down to $\mathcal{O}(10^{-16})$.

We can rewrite the conversion rate as
\begin{align}
\mathrm{Cr}(\mu\text{--}e)
= \frac{\alpha_s^2}{4 m_S^4 \Gamma_{\mathrm{capture}}}
y_L^{11} y_R^{11} y_L^{12} y_R^{12} f(r),
\end{align}
where
\begin{align}
f(r) &= \frac{\left[r^{12} - (\alpha_v/\alpha_s)\right]^2}{r^{11} r^{12}}
+ r^{11} r^{12} \left( \frac{1}{r^{12}} - \frac{\alpha_v}{\alpha_s} \right)^2,
\label{f(r)}
\end{align}
with $r^{1i} = y_L^{1i}/y_R^{1i}$. We then obtain the constraint
\begin{align}
y_L^{11} y_R^{11}y_L^{12} y_R^{12}
< \mathrm{Cr}(\mu\text{--}e)
\frac{1}{f(r)}\frac{4 m_S^4 \Gamma_{\mathrm{capture}}}{\alpha_s^2}.
\label{yLR1112Cond}
\end{align}

\begin{figure}[ht!]
	\includegraphics[width=0.50\linewidth]{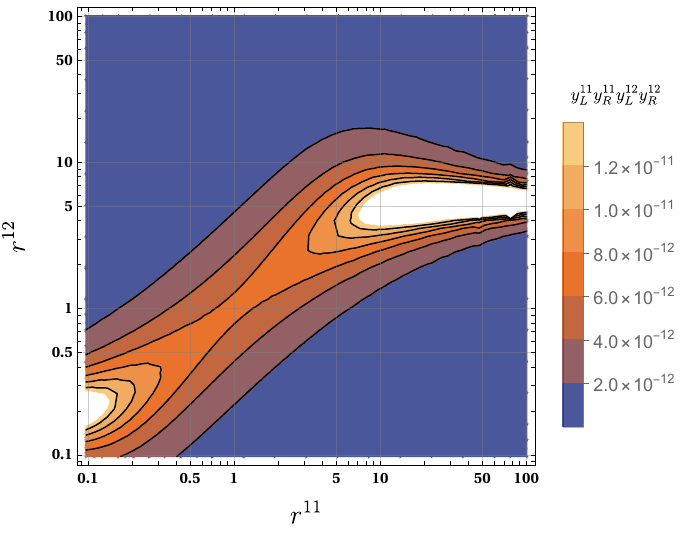}
	\includegraphics[width=0.49\linewidth]{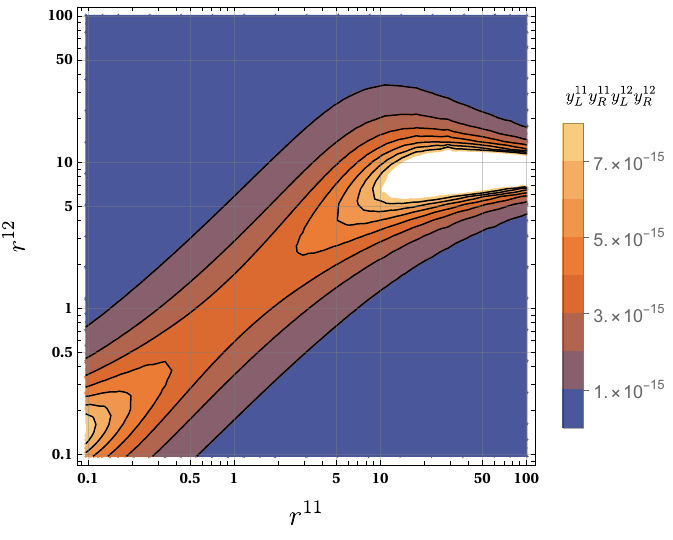}
\caption{Allowed parameter regions for $\mu\text{--}e$ conversion in gold (left) and aluminum (right) nuclei.}
	\label{mu2e}
\end{figure}

In Fig.~\ref{mu2e}, we show the dependence of the coupling product $y_L^{11} y_R^{11} y_L^{12} y_R^{12}$ on the $(r^{11}, r^{12})$ parameter plane. The corresponding limits for gold and aluminum nuclei are displayed in the left and right panels of Fig.~\ref{mu2e}, respectively, using the current SINDRUM II bound. Notably, in the special cases where $r^{12} = \alpha_v/\alpha_s$ or $r^{12} = \alpha_s/\alpha_v$, one of the two terms in Eq.~(\ref{f(r)}) vanishes. If, in addition, $r^{11}$ becomes either very small or very large, the function $f(r)$ is strongly suppressed. In this regime, the constraint is significantly relaxed, allowing the coupling product to reach values as large as $\mathcal{O}(10^{-11})$ for Au nuclei and $\mathcal{O}(10^{-15})$ for Al nuclei. In more general cases, $r^{11}$ and $r^{12}$ can differ by about one order of magnitude, e.g., $r^{11} \simeq 10\times r^{12}$, while still preserving a normal hierarchical structure of the couplings.

\section{\label{sec6}Singlet leptoquark production at proton--proton colliders}
We consider pair production of the singlet scalar leptoquark $S$ at proton--proton colliders,
\begin{equation}
pp \to S S^\dagger,
\end{equation}
which proceeds dominantly via QCD interactions through gluon fusion and quark--antiquark annihilation. The corresponding partonic cross sections are given at leading order by~\cite{Kramer:1997hh}
\begin{align}
\hat\sigma_{q\bar q}(\hat{s}) &=\frac{2\pi \alpha_s^2}{27\hat s}\,\beta^3,\\
\hat\sigma_{gg}(\hat{s})&=\frac{\pi \alpha_s^2}{96\hat s}
\left[\beta (41 - 31\beta^2)+
(18\beta^2 - \beta^4 -17)\ln\frac{1+\beta}{1-\beta}
\right],
\end{align}
where $\beta = \sqrt{1 - 4m_S^2/\hat{s}}$. The hadronic cross section is obtained via convolution with parton distribution functions,
\begin{equation}
\sigma(pp \to SS^\dagger)=\int_{\tau_0}^{1} d\tau \,
\Big[\mathcal{L}_{gg}(\tau,\mu^2)\,\hat\sigma_{gg}(\hat{s})
+\mathcal{L}_{q\bar q}(\tau,\mu^2)\,\hat\sigma_{q\bar q}(\hat{s})\Big],
\end{equation}
where $\tau = \hat{s}/s$, $\tau_0 = 4m_S^2/s$, and $\mu =\mathcal{O}(m_S)$. The parton luminosities follow the standard definitions~\cite{Martin:2009iq},
\begin{align}
\mathcal{L}_{gg}(\tau,\mu^2)&=\int_\tau^1 \frac{dx}{x}\,
g(x,\mu^2)\,g\!\left(\frac{\tau}{x},\mu^2\right),\\
\mathcal{L}_{q\bar q}(\tau,\mu^2)&=\sum_q\int_\tau^1 \frac{dx}{x}\,
\Big[q(x,\mu^2)\bar q\!\left(\frac{\tau}{x},\mu^2\right)+
\bar q(x,\mu^2)q\!\left(\frac{\tau}{x},\mu^2\right)\Big].
\end{align}

The singlet leptoquark decays into up-type quarks and charged leptons via both left- and right-handed Yukawa interactions. Due to the different flavor structures of $\tilde y_L^{ia}$ and $y_R^{ab}$, the corresponding partial widths are
\begin{align}
\Gamma(S \to u_i \ell_a)
&=
\frac{m_S}{16\pi}|\tilde y_L^{ia}|^2,
\\
\Gamma(S \to u_a \ell_b)
&=
\frac{m_S}{16\pi}|y_R^{ab}|^2,
\end{align}
while the decay into down-type quarks is given by
\begin{align}
\Gamma(S \to d_i \nu_a)
=
\frac{m_S}{16\pi}|\tilde y_L^{ia}|^2.
\end{align}
The branching ratio into the phenomenologically relevant mode $S \to t\mu$ is therefore
\be \mathrm{Br}(S \to t\mu) = \frac{|\tilde y_L^{32}|^2 + |y_R^{32}|^2}{2\sum_{i,a}|\tilde y_L^{ia}|^2 +\sum_{a,b} |y_R^{ab}|^2}.\ee

The Yukawa structure is constrained by low-energy observables, including $\Delta a_\mu$ and charged lepton flavor violation, which strongly suppress first-generation couplings while allowing sizable third-generation interactions. To incorporate these effects, we perform a scan over the allowed parameter space and define an effective branching ratio $\mathrm{Br}_{\rm eff}(S \to t\mu)$, typically of order $\mathrm{Br}_{\rm eff} \sim \mathcal{O}(0.2 - 1)$. Under the narrow-width approximation, valid for $\Gamma_S \ll m_S$, the observable signal cross section is
\begin{equation}
\sigma_{\rm sig}(pp \to t\bar t\,\mu^+\mu^-)=\sigma(pp \to SS^\dagger)\,\mathrm{Br}_{\rm eff}^2.
\end{equation}

Numerically, we evaluate the cross section at $\sqrt{s} = 14~\mathrm{TeV}$ using leading-order PDFs with scale choice $\mu = m_S/2$, and include a $K$-factor $K \simeq 2$. The production cross section decreases rapidly with increasing $m_S$,
\begin{equation}
\sigma(pp \to SS^\dagger) \sim
\begin{cases}
\mathcal{O}(10)\ \mathrm{fb} & m_S \sim 1~\mathrm{TeV}, \\
\mathcal{O}(10^{-1})\ \mathrm{fb} & m_S \sim 2~\mathrm{TeV}, \\
\mathcal{O}(10^{-4})\ \mathrm{fb} & m_S \sim 3~\mathrm{TeV}.
\end{cases}
\end{equation}
This behavior is mainly driven by the steep fall-off of parton distribution functions at large momentum fractions and by threshold suppression. The expected event number at the HL-LHC ($\mathcal{L}=3000~\mathrm{fb}^{-1}$), computed using $\sigma_{\rm sig} = \sigma \times \mathrm{Br}_{\rm eff}^2$ with a representative value $\mathrm{Br}_{\rm eff}=0.5$, ranges from $\mathcal{O}(10^{4})$ down to $\mathcal{O}(1\text{--}10)$ for $m_S = 1$--$3~\mathrm{TeV}$.

Current LHC searches constrain $m_S \gtrsim 1.5$--$2~\mathrm{TeV}$~\cite{CMS:2020wzx, ATLAS:2020xov}, assuming dominant third-generation final states. In our framework, the muon $g-2$ requirement favors heavier masses, pushing the viable region to the multi-TeV regime where direct production becomes suppressed. Future hadron colliders operating at $\sqrt{s} = 27~\mathrm{TeV}$ can significantly extend the sensitivity to multi-TeV leptoquarks.

\section{\label{sec7}Conclusion}
In this work, we have extended the 331LHN model by introducing scalar leptoquarks. The quark and lepton structure of the model accommodates a rich set of leptoquark multiplets, including singlet, triplet, sextet, and octet representations. We have identified their particle content and constructed the corresponding Yukawa interactions with the fermions of the model.

Focusing on the singlet scalar leptoquark, we have shown that it can generate sizable chirality-enhanced contributions to the muon anomalous magnetic moment. In particular, it can account for the $4.2\sigma$ discrepancy reported by the 2021 Muon $g-2$ experiment. Taking into account the updated 2025 sensitivity, we find that the allowed parameter space favors a relatively heavy leptoquark with mass $m_S \gtrsim \mathcal{O}(\text{few})~\mathrm{TeV}$, with current bounds already requiring $m_S \gtrsim 1.8~\mathrm{TeV}$.

We have further analyzed the combined constraints from $\Delta a_\mu$ and charged lepton flavor violation. The singlet leptoquark induces $\mu$--$e$ conversion at tree level, making CLFV observables particularly powerful probes of the model. Using the latest bounds, we obtain strong constraints on the Yukawa couplings. In particular, the combination $y_L^{32} y_R^{32}$ is constrained to be below $2.54 \times 10^{-3}$, while first-generation couplings are suppressed to the level of $10^{-6}$ and third-generation couplings remain at most $\mathcal{O}(10^{-2})$. The combination relevant for $\mu$--$e$ conversion satisfies $y_L^{11} y_R^{11} y_L^{12} y_R^{12} \leq 1.2 \times 10^{-11}$.

We have also investigated the collider phenomenology of the singlet leptoquark. Its production at hadron colliders is dominated by QCD-induced pair production, which depends primarily on its mass. However, the observable signal rates are controlled by the Yukawa couplings through the decay branching ratios. Taking into account the constraints from low-energy observables, the viable parameter space typically leads to suppressed production rates at the LHC for multi-TeV masses. As a result, direct detection at the current LHC is challenging, while future high-energy colliders offer significantly improved sensitivity.

Future experimental prospects will play a crucial role in testing this scenario. Next-generation CLFV experiments such as Mu2e and COMET are expected to significantly improve the sensitivity to $\mu$--$e$ conversion, probing regions of parameter space that are currently inaccessible. At the same time, improved measurements of the muon anomalous magnetic moment, together with searches for leptoquarks at the high-luminosity LHC and future hadron colliders, will provide complementary probes. The interplay of these measurements will be essential to either confirm this framework or further constrain the allowed parameter space.

\bibliographystyle{JHEP}

\bibliography{LQ-331}

\end{document}